\documentclass[journal]{IEEEtran}
\usepackage{amsmath,latexsym,amssymb,stmaryrd}
\usepackage{setspace,stackrel,tikz,graphicx}
\usepackage{exscale,relsize,subfig,textcomp,stackrel,setspace,float}
\usepackage{mdframed,pifont}
\usepackage{multicol,xfrac}
\usepackage{multirow, booktabs}
\usepackage{mathtools}
\usepackage[T1]{fontenc}
\usepackage[font=footnotesize]{caption}
\usepackage{lipsum}
\usepackage{multicol}
\usepackage{graphicx}
\usepackage[utf8]{inputenc}

\usepackage{amsthm}
\definecolor{blue}{rgb}{0.16, 0.32, 0.75}
\begin{document}

\title{\vspace{0cm}\LARGE Dual-Mode Chirp Spread Spectrum Modulation\vspace{0em}}
\makeatletter
\patchcmd{\@maketitle}
  {\addvspace{0\baselineskip}\egroup}
  {\addvspace{0\base
  {}lineskip}\egroup}
  {}
  \makeatother
\author{Ali Waqar Azim,  Ahmad Bazzi, Raed Shubair, Marwa Chafii
\thanks{Ali Waqar Azim is with Department of Telecommunication Engineering,  University of Engineering and Technology,  Taxila,  Pakistan (email: aliwaqarazim@gmail.com).}
\thanks{Ahmad Bazzi and Raed Shubair is with with Engineering Division, New York University (NYU) Abu Dhabi, 129188, UAE (email: \{ahmad.bazzi,raed.shubair\}@nyu.edu).}
\thanks{Marwa Chafii is with Engineering Division, New York University (NYU) Abu Dhabi, 129188, UAE and NYU WIRELESS, NYU Tandon School of Engineering, Brooklyn, 11201, NY, USA (email: marwa.chafii@nyu.edu).}}
\maketitle
\begin{abstract}
In this letter, we propose dual-mode chirp spread spectrum (DM-CSS) modulation for low-power wide-area networks. DM-CSS is capable of achieving a higher spectral efficiency (SE) relative to its counterparts, such as Long Range (LoRa) modulation. Considering the same symbol period, the SE in DM-CSS are augmented by: (i) simultaneously multiplexing even and odd chirp signals; (ii) using phase shifts of \(0\) and \(\pi\) radians for both even and odd chirp signals; and (iii) using either up-chirp or down-chirp signal. The SE increases by up to \(116.66\%\) for the same bandwidth and spreading factor relative to LoRa. We present a complete transceiver architecture along with non-coherent detection process. Simulation results reveal that DM-CSS is not only more spectral efficient but also more energy efficient than most classical counterparts. It is also demonstrated that DM-CSS is robust to phase and frequency offsets. 
\end{abstract}
\begin{IEEEkeywords}
LoRa, Low power wide area networks, chirp spread spectrum, IoT.
\end{IEEEkeywords}
\IEEEpeerreviewmaketitle

\section{Introduction}
\IEEEPARstart{L}{ong} Range (LoRa) is a frequency-shift chirp spread spectrum (CSS) modulation \cite{vangelista2017frequency}, in which, the number of bits transmitted per symbol is equal to the spreading factor, \(\lambda\), that results in \(N= 2^\lambda\) chirp symbols, where \(N\) is the number of cyclic time-shifts of the chirp signal. Despite the fact that LoRa offers a wide range of throughputs, and can target a multitude of applications, one of the limiting factors is the low achievable bit rate/spectral efficiency (SE). In this letter, we define SE as the achievable bit rate in a given bandwidth, and energy efficiency (EE) as the required signal-to-noise ratio (SNR) for correct bit detection at a given bit-error rate (BER).

Several studies have proposed energy and/or spectral efficient variants of LoRa. Some alternatives are \cite{bomfin2019novel,epsk_lora,hanif2020slope,elshabrawy2019interleaved,ssk_ics_lora,gcss,dcrk_css}. In \cite{bomfin2019novel}, the authors propose phase-shift keying-LoRa (PSK-LoRa), that encodes  additional bits in the phase shifts (PSs) of the chirp signal, where quaternary PSs result in the optimal performance in terms of EE. In addition, \(\lambda + 2\) bits per symbol are transmitted. The performance of PSK-LoRa can be further improved with enhanced PSK-LoRa (ePSK-LoRa) \cite{epsk_lora}. The optimal variant of ePSK-LoRa transmits \(\lambda + 3\) bits per symbol. Nonetheless, both PSK-LoRa and ePSK-LoRa require coherent detection to determine the information embedded in the PSs, and are capable of transmitting only \(2\) or \(3\) additional bits relative to LoRa, respectively. Another approach that transmits \(\lambda+1\) bits per symbol by extending the signal space using negative frequency change slope is slope-shift keying-LoRa (SSK-LoRa) \cite{hanif2020slope}. In interleaved chirp spreading LoRa (ICS-LoRa) \cite{elshabrawy2019interleaved}, the use of interleaved LoRa symbols provide an additional dimension to transmit \(\lambda +1\) bits per symbol. Both SSK-LoRa and ICS-LoRa only marginally increase the SE relative to LoRa because only \(1\) additional bit is transmitted per symbol in these approaches compared to LoRa. SSK-ICS-LoRa \cite{ssk_ics_lora} amalgamates SSK-LoRa and ICS-LoRa and uses up chirps, down chirps (for signal spreading), and interleaved versions of both, enabling it to transmit \(\lambda + 2\) bits per symbol. Though SSK-ICS-LoRa outperforms LoRa, SSK-LoRa and ICS-LoRa in terms of the EE and SE, it  only increases the number of transmitted bits per symbol by \(2\) relative to LoRa: that is unsubstantial. Recently, group-based CSS (GCSS) has been proposed in \cite{gcss}, that is configurable via so-called group number, \(\mathrm{G}\), which provides a trade-off between EE and SE. By increasing \(\mathrm{G}\), the EE decreases, but the SE increases and vice versa. The number of transmitted bits per symbol for GCSS using \(\mathrm{G}=2\) is \(2\lambda-2\). Note that while all the variants of GCSS are more spectral efficient than LoRa, only a few are energy efficient because increasing \(\mathrm{G}\) reduces the EE. The scheme in \cite{dcrk_css}, i.e., discrete chirp-rate keying (DCRK)-CSS, transmits additional bits in the discrete chirp rate. Considering \(M_\mathrm{c}\) chirp rates, the number of transmitted bits per symbol is \(\lambda + \log_2(M_\mathrm{c})\). DCRK-CSS improves both the EE and SE relative to LoRa, however, an increase in \(M_\mathrm{c}\) considerably increases the complexity of DCRK-CSS receiver, and the increase in the number of bits is limited by a factor of \(\log_2(M_\mathrm{c})\).

In this letter, we propose dual-mode CSS (DM-CSS) modulation as an alternative to the classical schemes as it circumvents most of their limitations. Unlike PSK-LoRa and ePSK-LoRa, both coherent and non-coherent detection is possible for DM-CSS. Moreover, contrary to the remainder of previously discussed schemes, DM-CSS almost doubles the number of transmitted bits per symbol relative to LoRa. In DM-CSS, which is an improved variant of the scheme in \cite{8937880}, \(N\) frequencies of an un-chirped symbol are divided into \(\sfrac{N}{2}\) even and \(\sfrac{N}{2}\) odd frequencies. Next, binary PSs are added to both the activated even and odd frequencies. Lastly, the un-chirped symbol is chirped using either the up-chirp or the down-chirp. It is highlighted that the term dual-mode refers to the use of both the even and the odd frequencies. It shall become evident in the sequel that the proposed DM-CSS has the following concrete advantages of the classical counterparts: (i) higher achievable SE, i.e.,  it is capable of transmitting higher number of bits per symbol for the same symbol duration and bandwidth; (ii) it is more energy efficient, in a sense that for the same throughput, less SNR/bit is needed; and (iii) both coherent and non-coherent detection is applicable.

We can summarize the contributions of this work as follows:
\begin{enumerate}
\item A novel DM-CSS modulation is proposed that achieves higher SE and better EE than other alternatives. 
\item A complete transceiver architecture for DM-CSS employing non-coherent detection is provided.
\item A framework for evaluating the orthogonality of DM-CSS symbols is provided. 
\item The performance of DM-CSS in terms of BER considering additive white Gaussian noise (AWGN) channel, fading channel, and in the presence of phase and frequency offsets is evaluated.
\end{enumerate}

The rest of the article is organized as follows. Section II provides the transceiver architecture and evaluates the orthogonality of the DM-CSS symbols. Section III provides simulation results and Section IV renders conclusions. 
\vspace{-2mm}
\section{Dual-Mode Chirp Spread Spectrum (DM-CSS)}
In this section, we introduce DM-CSS modulation. Table \ref{table1} provides a summary of different LoRa variant time-domain signal structures, achievable SE in bits/s/Hz, and relative increase in SE over LoRa in bits/s/Hz.

From Table \ref{table1}, it can be observed that the information bearing element in LoRa is cyclic time-shift of the chirp signal, i.e., \(k\). In DM-CSS, besides the information being transmitted in use of either up-chirp or down-chirp, it is also transmitted via the activated even and odd activated frequencies and their corresponding PSs, i.e., \(k_\mathrm{e}, k_\mathrm{o}, \alpha_\mathrm{e}\) and \(\alpha_\mathrm{o}\). ePSK-LoRa transmits information in the PSs of \(l\) sub-blocks, i.e., \(p_l\), where \(l=\llbracket 1,N_\mathrm{b} \rrbracket\), and the fundamental frequency, \(k\) \cite{epsk_lora}. SSK-ICS-LoRa uses up chirps, down chirps, interleaved up-chirp and interleaved down-chirp. In GCSS, SE depends on \(\mathrm{G}\), where an increase in \(\mathrm{G}\) results in higher achievable SE. DCRK-CSS uses different \(M_\mathrm{c}\) to increase the number of bits transmitted per symbol.
\renewcommand{\arraystretch}{1.2}
\begin{table*}[h]
  \centering
   \caption{Characteristics of Different LoRa variants. Here, \(\lambda = log_2(N)\), \(N_\mathrm{b}\) and \(M_\phi\) are the number of sub-blocks and the number of PSs used in ePSK-LoRa, \(\Pi[\cdot]\) refers to interleaved version of a chirp symbol. }
   \label{table1}
   \resizebox{\textwidth}{!}{%
  \begin{tabular}{@{}ccccc@{}}
    \toprule
       \toprule
\bfseries {Modulation}    & \bfseries {Transmit Symbol} & \bfseries {SE} & \bfseries {SE increase w.r.t. LoRa} \\
    \midrule
LoRa & \(\exp\left\{j\frac{2\pi}{N}n(k+n)\right\}\)&\(\frac{\lambda}{N}\) & \(-\) \\
DM-CSS &\(\alpha_\mathrm{e}\exp\left\{j\frac{2\pi}{N}n \left(k_\mathrm{e} \pm n\right)\right\}+\alpha_\mathrm{o}\exp\left\{j\frac{2\pi}{N}n \left(k_\mathrm{o} \pm n\right)\right\}\) &\(\frac{2\lambda +1}{N}\) &\(\frac{\lambda+1}{N}\)\\
ePSK-LoRa &\(\sum_{l=1}^{N_\mathrm{b}}\exp\left\{j2\pi\left(\frac{kn}{N}+\frac{ln}{N_\mathrm{b}}+\frac{p_l}{2^{\eta_\mathrm{PS}}}+\frac{n^2}{N}\right)\right\}\) &\(\frac{\log_2(\sfrac{N}{N_\mathrm{b}}) +N_\mathrm{b}\log_2(M_\phi)}{N}\)  & \(\frac{3}{N}\)  \scriptsize{(for \(\{N_\mathrm{b},M_\phi\}=\{2,4\}\))}\\
SSK-ICS-LoRa & \(\exp\left\{j\frac{2\pi}{N}n(k \pm n)\right\}\), \(\Pi\left[\exp\left\{j\frac{2\pi}{N}n(k \pm n)\right\}\right]\)& \(\frac{\lambda +2}{M}\) & \(\frac{2}{N}\)\\
 GCSS & \(\sum_{i=1}^{\mathrm{G}}\exp\left\{j\frac{2\pi}{N}n(k_i+n)\right\}\), \(k_i= \llbracket (i-1)\frac{N}{\mathrm{G}},i\left(\frac{N}{\mathrm{G}}-1\right) \rrbracket\)&\(\frac{\mathrm{G}\log_2(\sfrac{N}{\mathrm{G}})}{N}\) &  \(\frac{\mathrm{G}\log_2(\sfrac{N}{G})-1}{N}\) \\
DCRK-CSS & \(\exp\left\{j\frac{\pi}{N}n(2k+M_\mathrm{c}n)\right\}\) &\(\frac{\lambda+\log_2(M_\mathrm{c})}{N}\) & \(\frac{\log_2(M_\mathrm{c})}{N}\) \\
    \bottomrule
      \bottomrule
  \end{tabular}}
\end{table*}
\vspace{-3.8mm}
\subsection{Transmission}
Consider that \(N\) frequencies are available in bandwidth \(B = \sfrac{N}{T_\mathrm{s}}\), where \(T_\mathrm{s}\) is the symbol period. Among these \(N\) frequencies, there are \(\sfrac{N}{2}\) even and \(\sfrac{N}{2}\) odd frequencies. The even frequencies and odd frequencies are denoted by indexes \(2\tilde{k}_\mathrm{e}\) and \(2\tilde{k}_\mathrm{o}+1\) for \(\tilde{k}_\mathrm{e} = \tilde{k}_\mathrm{o} = \llbracket 0, \sfrac{N}{2}-1\rrbracket\), respectively. One even, \(k_\mathrm{e}\), and one odd frequency, \(k_\mathrm{o}\) is activated among the possible \(\sfrac{N}{2}\) even and \(\sfrac{N}{2}\) odd frequencies. The activated frequencies (even and odd) have a PS of either \(0\) or \(\pi\) radians: which is denoted by \(\alpha\). If the PS is \(0\) radians, then \(\alpha = 1\), on the contrary, if the PS is \(\pi\) radians, then \(\alpha = -1\). The discrete representation of the un-chirped symbol is given as:
\begin{equation}\label{eq1}
f(n) = \alpha_\mathrm{e}\exp\left\{j\frac{2\pi}{N}k_\mathrm{e}n\right\}+\alpha_\mathrm{o}\exp\left\{j\frac{2\pi}{N}k_\mathrm{o}n\right\},
\end{equation}
for \(n = \llbracket 0, N-1\rrbracket\), where \(j = \sqrt{-1}\). \(\alpha_\mathrm{e}\) and \(\alpha_\mathrm{o}\) represents the PSs for even and odd frequencies, respectively. \(f(n)\) is then chirped using either an up-chirp, \(c_\mathrm{u}(n)=\exp\left\{j\frac{2\pi}{N}n^2\right\}\), or a down-chirp, \(c_\mathrm{d}(n)=\exp\left\{-j\frac{2\pi}{N}n^2\right\}\). The chirped signal, \(s(n) =f(n)c_{\mathrm{u/d}}(n)\) is given as:
\begin{equation}\label{eq2}
s(n)=\alpha_\mathrm{e}\exp\left\{j\frac{2\pi}{N}n \left(k_\mathrm{e} \pm n\right)\right\}+\alpha_\mathrm{o}\exp\left\{j\frac{2\pi}{N}n \left(k_\mathrm{o} \pm n\right)\right\},
\end{equation}
where \(c_{\mathrm{u/d}}(n)\) refers to either an up-chirp or a down-chirp.

DM-CSS can be extended to $p-$mode type encoding, however, the resulting design may not be energy efficient and it will come at a cost of higher complexity.The symbol energy of \(s(n)\) is \(E_\mathrm{s}=\sfrac{1}{N}\sum_{n=0}^{N-1}\vert s(n)\vert^2=2\).

The breakdown of the number of bits encoded in the DM-CSS symbol is as follows: (i) \(\lambda_\mathrm{e} = \log_2(\sfrac{N}{2})\) and \(\lambda_\mathrm{o} = \log_2(\sfrac{N}{2})\) bits are encoded in \(k_\mathrm{e}\) and \(k_\mathrm{o}\), respectively; (ii) \(\lambda_\mathrm{PS}=2\log_2(2)=2\) bits are encoded in \(\alpha_\mathrm{e}\) and \(\alpha_\mathrm{o}\); and (iii) \(\lambda_\mathrm{c}=\log_2(2)=1\) bit is encoded in the use of either up-chirp or the down-chirp. Thus, the total number of bits that can be transmitted per DM-CSS symbol of duration \(T_\mathrm{s}\) is
\begin{equation}\label{eq4}
\lambda_\mathrm{DM-CSS} = \lambda_\mathrm{e} + \lambda_\mathrm{o} +\lambda_\mathrm{PS} + \lambda_\mathrm{c} = 2\lambda + 1.
\end{equation}

Given that the bit rate for DM-CSS is \(R = \sfrac{\lambda_\mathrm{DM-CSS}}{T_\mathrm{s}}\), and \(B=\sfrac{M}{T_\mathrm{s}}\), the SE for DM-CSS, \(\eta_\mathrm{DM-CSS}\) is given as:
\begin{equation}\label{se}
\eta_\mathrm{DM-CSS} = \frac{R}{B}= \frac{2\lambda + 1}{N}.
\end{equation}

From (\ref{se}), it is evident that the relative SE increase for DM-CSS over LoRa is \(\sfrac{(\lambda+1)}{N}\), which yields a SE increase of \(116.66\%\) and \(108.33\%\) over LoRa for \(\lambda = 6\) and \(\lambda = 12\), respectively.
\vspace{-4mm}
\subsection{Detection}
Here we only provide the non-coherent detection process which does not require any a priori knowledge of the channel because of its practicality as the coherent detector is generally very complex. However, to counterbalance any phase rotation of the channel (if it is \(\geq \sfrac{\pi}{2}\)), a coherent detector that has a priori knowledge of the channel is mandatory.

Firstly, the received symbol, \(r(n)\) is multiplied with both an up-chirp and a down-chirp yielding \(r_1(n) = r(n)c_\mathrm{d}(n)\) and \(r_2(n) = r(n)c_\mathrm{u}(n)\). Next, \(R_1(k)\) and \(R_2(k)\) are obtained by evaluating the discrete Fourier transform (DFT) for \(r_1(n)\) and \(r_2(n)\), respectively. By using  \(\kappa_1 =\max\left\{\vert R_1(k)\vert\right\}\), and \(\kappa_2 = \max\left\{\vert R_2(k)\vert\right\}\) for \(k=\llbracket 0,N-1\rrbracket\), the following criterion:
\begin{equation}\label{eq5}
\kappa_1 \lessgtr \kappa_2.
\end{equation}
determines whether an up-chirp was used or a down-chirp was used for chirping at the transmitter. If \(\kappa_1 > \kappa_2\), then an up-chirp was used, conversely if \(\kappa_2>\kappa_1\), a down-chirp was used. 

Using non-coherent detection process, the activated even and odd frequencies are estimated as:
\begin{equation}\label{eqa1}
\hat{k}_\mathrm{e}= \mathrm{arg}\max_{k \in 2\tilde{k}_\mathrm{e}}~\left \vert R(k) \right\vert,~~\text{and}~~\hat{k}_\mathrm{o}= \mathrm{arg}\max_{k \in 2\tilde{k}_\mathrm{o}+1}~\left \vert R(k) \right\vert,
\end{equation}
respectively, where
\begin{equation}\label{Rk}
R(k)= \left\{ 
  \begin{array}{ c l }
    R_1(k) & \quad \kappa_1 > \kappa_2\\
  R_2(k)& \quad \kappa_1 < \kappa_2
   \end{array}
\right..
\end{equation}

The PSs of even and odd frequencies are respectively evaluated by determining the polarity of \(R(\hat{k}_\mathrm{e})\) and \(R(\hat{k}_\mathrm{o})\) as:
\begin{equation}\label{eqa3}
\hat{\alpha}_\mathrm{e}= \mathrm{polarity}\left\{R(\hat{k}_\mathrm{e})\right\},~~\text{and}~~\hat{\alpha}_\mathrm{o}= \mathrm{polarity}\left\{R(\hat{k}_\mathrm{o})\right\},
\end{equation}
where \(\mathrm{polarity}\{\cdot\}\) is the maximum likelihood criteria to determine the polarity of the input frequency index. If the output of \(\mathrm{polarity}\{\cdot\} > 0\), then \(\hat{\alpha}_\mathrm{e}/\hat{\alpha}_\mathrm{o} = 1\), conversely, \(\hat{\alpha}_\mathrm{e}/\hat{\alpha}_\mathrm{o} = -1\). It is accentuated that if the channel's phase rotation is \(\geq \sfrac{\pi}{2}\), then, only coherent detection would be viable because the \(\mathrm{polarity}\{\cdot\}\) function will no longer be applicable.
\subsection{Orthogonality of DM-CSS Symbols}
To establish the orthogonality of DM-CSS symbol, we evaluate the inner product of two distinct symbols as \(\langle s(n),\tilde{s}(n)\rangle= \sum_{n=0}^{N-1} s(n) \overline{\tilde{s}}(n)\). Note that the activated even and odd frequencies and PSs in \(s(n)\) are \(k_\mathrm{e}\), \(k_\mathrm{o}\), \(\alpha_\mathrm{e}\), and \(\alpha_\mathrm{o}\), whereas the activated even and odd frequencies and PS in \(\tilde{s}(n)\) are \(\tilde{k}_\mathrm{e}\), \(\tilde{k}_\mathrm{o}\), \(\tilde{\alpha}_\mathrm{e}\), and \(\tilde{\alpha}_\mathrm{o}\). Note that the following conditions must hold to evaluate the orthogonality: (i) \(k_\mathrm{e} \neq \tilde{k}_\mathrm{e}\); and  (ii) \(k_\mathrm{o} \neq \tilde{k}_\mathrm{o}\). Moreover, it shall become evident that the PSs and chirps do not impact the orthogonality of the symbols, therefore, it does not matter if they are same or different. Additionally, there are four distinct cases for which the orthogonality should be evaluated. The cases are: (i) when both \(s(n)\) and \(\tilde{s}(n)\) are attained using an up-chirp, i.e., \(s(n)=f(n)c_\mathrm{u}(n)\) and \(\tilde{s}(n)=\tilde{f}(n)c_\mathrm{u}(n)\); (ii) when both \(s(n)\) and \(\tilde{s}(n)\) are attained using a down-chirp, i.e., \(s(n)=f(n)c_\mathrm{d}(n)\) and \(\tilde{s}(n)=\tilde{f}(n)c_\mathrm{d}(n)\); (iii) when \(s(n)\) is attained using an up-chirp and \(\tilde{s}(n)\) is attained using a down-chirp, i.e., \(s(n)=f(n)c_\mathrm{u}(n)\) and \(\tilde{s}(n)=\tilde{f}(n)c_\mathrm{d}(n)\); and (iv) when \(s(n)\) is attained using a down-chirp and \(\tilde{s}(n)\) is attained using an up-chirp, i.e., \(s(n)=f(n)c_\mathrm{d}(n)\) and \(\tilde{s}(n)=\tilde{f}(n)c_\mathrm{u}(n)\). In the sequel, we consider these cases separately.
\subsubsection*{Case I: \(s(n)=f(n)c_\mathrm{u}(n)\) and \(\tilde{s}(n)=\tilde{f}(n)c_\mathrm{u}(n)\)}
In this case, \(\langle s(n),\tilde{s}(n)\rangle\) evaluates to 
\begin{equation}\label{eq7}
\begin{split}
\langle s(n),\tilde{s}(n)\rangle & = \alpha_\mathrm{e}\tilde{\alpha}_\mathrm{e} \underbrace{\sum_{n=0}^{N-1}\exp\left\{j\frac{2\pi}{N}n\left(k_\mathrm{e}-\tilde{k}_\mathrm{e}\right)\right\}}_{:=\tau_1}\\
&+\alpha_\mathrm{e}\tilde{\alpha}_\mathrm{o} \underbrace{\sum_{n=0}^{N-1}\exp\left\{j\frac{2\pi}{N}n\left(k_\mathrm{e}-\tilde{k}_\mathrm{o}\right)\right\}}_{:=\tau_2}\\
& +\alpha_\mathrm{o}\tilde{\alpha}_\mathrm{e} \underbrace{\sum_{n=0}^{N-1}\exp\left\{j\frac{2\pi}{N}n\left(k_\mathrm{o}-\tilde{k}_\mathrm{e}\right)\right\}}_{:=\tau_3}\\
&+\alpha_\mathrm{o}\tilde{\alpha}_\mathrm{o} \underbrace{\sum_{n=0}^{N-1}\exp\left\{j\frac{2\pi}{N}n\left(k_\mathrm{o}-\tilde{k}_\mathrm{o}\right)\right\}}_{:=\tau_4}.
\end{split}
\end{equation}

For \(k_\mathrm{e} \neq \tilde{k}_\mathrm{e}\)
\begin{equation}\label{eq8}
\begin{split}
\tau_1= \frac{1-\exp\left\{j2\pi (k_\mathrm{e}-\tilde{k}_\mathrm{e})\right\}}{1-\exp\left\{j\frac{2\pi}{N} (k_\mathrm{e}-\tilde{k}_\mathrm{e})\right\}}=0. 
\end{split}
\end{equation}

\(\tau_2=\tau_3 =\tau_4 = 0\), no matter what since the odd and even frequencies can never meet. Thus, \(\langle s(n),\tilde{s}(n)\rangle = 0\) when \(s(n)=f(n)c_\mathrm{u}(n)\) and \(\tilde{s}(n)=\tilde{f}(n)c_\mathrm{u}(n)\).
\subsubsection*{Case II: \(s(n)=f(n)c_\mathrm{d}(n)\) and \(\tilde{s}(n)=\tilde{f}(n)c_\mathrm{d}(n)\)} In this particular case, the inner product, \(\langle s(n),\tilde{s}(n)\rangle\) is equal to:
\begin{equation}\label{eq9}
\begin{split}
\langle s(n),\tilde{s}(n)\rangle & = \alpha_\mathrm{e}\tilde{\alpha}_\mathrm{e} \overline{\tau}_1+\alpha_\mathrm{e}\tilde{\alpha}_\mathrm{o}\overline{\tau}_2+\alpha_\mathrm{o}\tilde{\alpha}_\mathrm{e} \overline{\tau}_3+\alpha_\mathrm{o}\tilde{\alpha}_\mathrm{o} \overline{\tau}_4,
\end{split}
\end{equation}
where, \(\overline{(\cdot)}\) denotes the conjugate of a complex argument. 

Since, \(\tau_1= \tau_2 = \tau_3 = \tau_4 = 0\), therefore,  \(\overline{\tau}_1= \overline{\tau}_2 = \overline{\tau}_3 = \overline{\tau}_4= 0\), which leads to \(\langle s(n),\tilde{s}(n)\rangle=0\).
\subsubsection*{Case III: \(s(n)=f(n)c_\mathrm{u}(n)\) and \(\tilde{s}(n)=\tilde{f}(n)c_\mathrm{d}(n)\)}
In this case, we have 
\begin{equation}\label{eq10}
\begin{split}
\langle s(n),\tilde{s}(n)\rangle & = \alpha_\mathrm{e}\tilde{\alpha}_\mathrm{e} \underbrace{\sum_{n=0}^{N-1}\exp\left\{j\frac{2\pi}{N}n\left(\alpha_1-\beta_1\right)\right\}}_{:=\tau_5}\\
&+\alpha_\mathrm{e}\tilde{\alpha}_\mathrm{o} \underbrace{\sum_{n=0}^{N-1}\exp\left\{j\frac{2\pi}{N}n\left(\alpha_1-\beta_2\right)\right\}}_{:=\tau_6}\\
& +\alpha_\mathrm{o}\tilde{\alpha}_\mathrm{e} \underbrace{\sum_{n=0}^{N-1}\exp\left\{j\frac{2\pi}{N}n\left(\alpha_2 -\beta_1\right)\right\}}_{:=\tau_7}\\
&+\alpha_\mathrm{o}\tilde{\alpha}_\mathrm{o} \underbrace{\sum_{n=0}^{N-1}\exp\left\{j\frac{2\pi}{N}n\left(\alpha_2-\beta_2\right)\right\}}_{:=\tau_8},
\end{split}
\end{equation}
where \(\alpha_1 := k_\mathrm{e}+n\), \(\alpha_2:=k_\mathrm{o}+n \), \(\beta_1:=\tilde{k}_\mathrm{e}+n\) and \(\beta_2:=\tilde{k}_\mathrm{o}+n \). Using the same property as in (\ref{eq8}) leads to \(\tau_5=\tau_6=\tau_7=\tau_8=0\), therefore, \(\langle s(n),\tilde{s}(n)\rangle=0\).
\subsubsection*{Case IV: \(s(n)=f(n)c_\mathrm{d}(n)\) and \(\tilde{s}(n)=\tilde{f}(n)c_\mathrm{u}(n)\)}
In this case, we have 
\begin{equation}\label{eq11}
\begin{split}
\langle s(n),\tilde{s}(n)\rangle & = \alpha_\mathrm{e}\tilde{\alpha}_\mathrm{e} \overline{\tau}_5+\alpha_\mathrm{e}\tilde{\alpha}_\mathrm{o}\overline{\tau}_6+\alpha_\mathrm{o}\tilde{\alpha}_\mathrm{e} \overline{\tau}_7+\alpha_\mathrm{o}\tilde{\alpha}_\mathrm{o} \overline{\tau}_8=0.
\end{split}
\end{equation}

Thus, it can be concluded that the DM-CSS symbols are always orthogonal and the PSs do not impact the orthogonality of the symbols.
 \section{Comparison with Other LoRa Variants}
In this section, we evaluate and compare the performance of the DM-CSS against LoRa, ePSK-LoRa, SSK-ICS-LoRa, GCSS, and DCRK-CSS. We consider ePSK-LoRa\((2,4)\) because of its resilience against the phase offset (PO) and frequency offset (FO). GCSS with \(\mathrm{G}=2\) is optimal in terms of EE and resilience to FO and PO; therefore, GCSS with only two groups is considered. Moreover, DCRK-CSS with \(M_\mathrm{c}=8\) is considered because of complexity issues. The performance of these schemes is evaluated in terms of (i) SE requirement versus required SNR/bit; (ii) BER performance in AWGN channel and fading channels; (iii) BER performance in AWGN considering PO; and (v) BER performance in AWGN considering FO.
\subsection{Spectral Efficiency vs. Required SNR Per Bit}
\begin{figure}[tb]\centering
\includegraphics[trim={65 0 0 2},clip,scale=0.72]{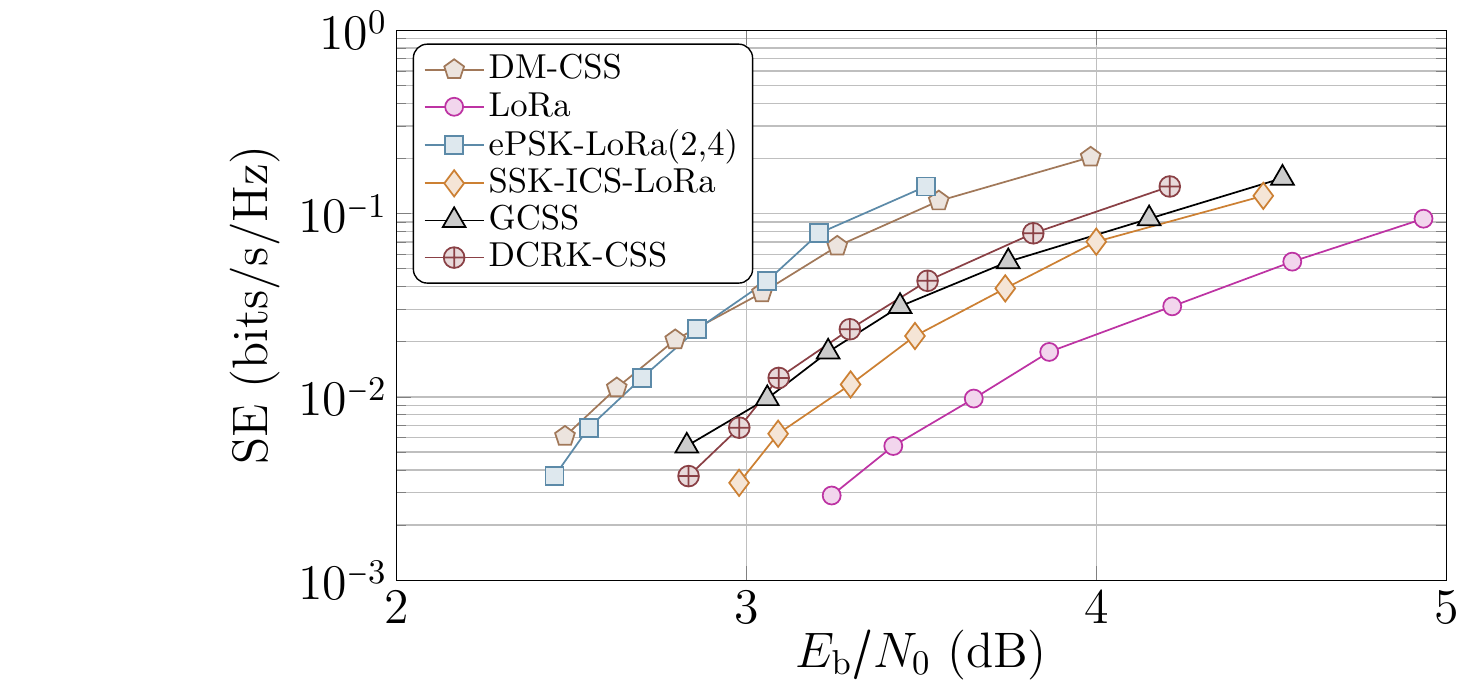}
  \caption{SE versus required SNR per bit for different schemes considering non-coherent detection and target BER of \(10^{-3}\) in AWGN channel.}
\label{fig1}
\end{figure}

Fig. \ref{fig1} compares the SE versus required SNR/bit performance of DM-CSS with other alternatives considering non-coherent detection in AWGN. SE is changed by varying \(\lambda = \llbracket 6,12\rrbracket\). Moreover, the EE is ascertained by evaluating the required \(E_\mathrm{b}/N_0=\sfrac{(E_\mathrm{s}T_\mathrm{s})}{(\eta_\mathrm{DM-CSS} N_0)}\) needed to attain BER of \(10^{-3}\) over \(10^6\) Monte Carlo runs. Here, \(N_0\) is the mono-lateral noise spectral density. The spectral efficiencies of DM-CSS and other counterparts are summarized in Table \ref{table1}. It is evident that DM-CSS has the highest achievable SE which clearly demonstrates its superiority in achieving maximum throughput.

From Fig.\ref{fig1}, we observe that DM-CSS provides the best trade-off between the achievable SE and required \(E_\mathrm{b}/N_0\). The modulation performance that coincides with that of DM-CSS is ePSK-LoRa(\(2,4\)). However, it is evident that DM-CSS provides a relative SE increase of \(\sfrac{(\lambda-2)}{M}\) bits/s/Hz \(\forall~\lambda\) over ePSK-LoRa(\(2,4\)). We can also observe that the performance of other alternatives, such as GCSS and DCRK-CSS are similar to each other and are better than SSK-ICS-LoRa. 
\subsection{Bit Error Rate Performance in AWGN and Fading Channels}
\begin{figure}[tb]\centering
\includegraphics[trim={52 0 0 0},clip,scale=0.86]{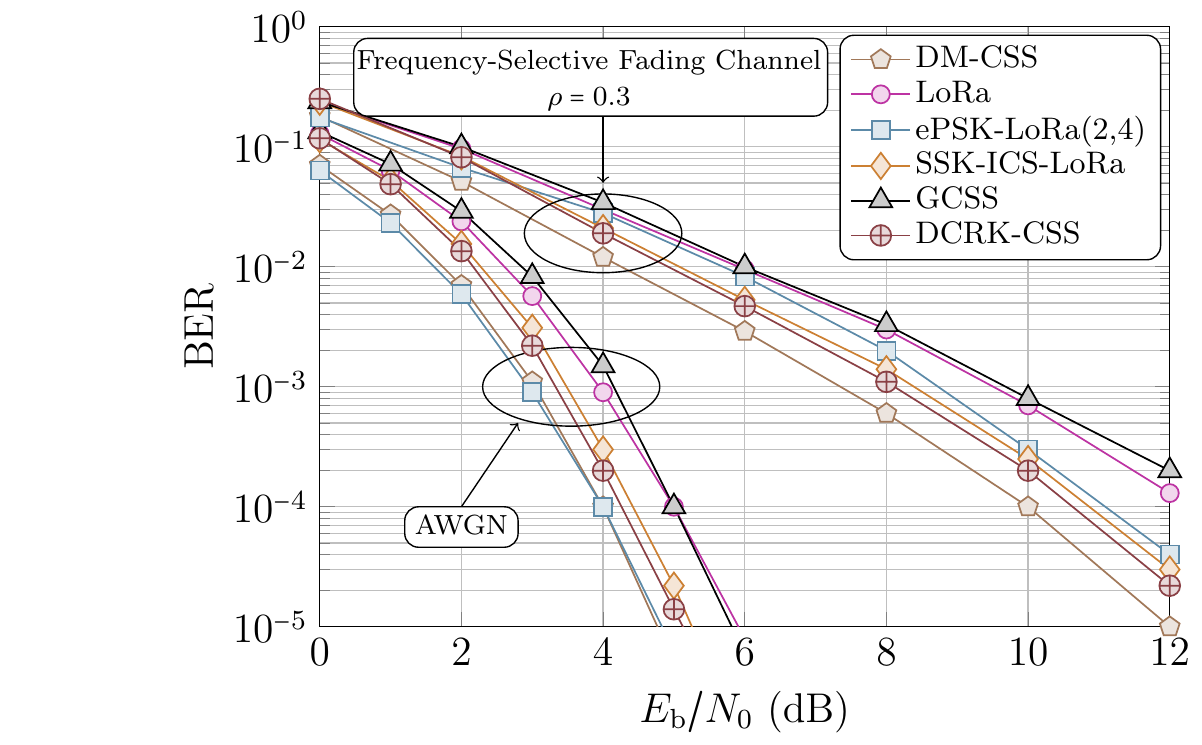}
  \caption{BER performance for different schemes considering non-coherent detection in AWGN channel for \(\lambda = 9\).}
\label{fig2}
\end{figure}

Next, we evaluate the performance of the discussed schemes considering an ideal linear AWGN and fading channels considering \(\lambda = 9\). These results are illustrated in Fig. \ref{fig2}. From Fig. \ref{fig2}, we can observe that in AWGN channel, the performance of DM-CSS is better than LoRa, SSK-ICS-LoRa, GCSS and DCRK-CSS, whereas, its performance is almost similar to that of ePSK-LoRa\((2,4)\). Thus, it is evident that one can use less energy to transmit more information with lower bit error probability when employing DM-CSS over other counterparts. We can observe that DM-CSS requires \(1\) dB less \(E_\mathrm{b}/N_0\) compared to LoRa at BER of \(10^{-3}\). On the other hand, for the same target BER, DM-CSS offers \(0.5\) dB improvement over GCSS and SSK-ICS-LoRa, and its gain over DCRK-CSS is around \(0.3\) dB.

We consider a frequency-selective \(2\)-tap fading channel having an impulse response of \(h(n) = \sqrt{1-\rho}\delta (nT) + \sqrt{\rho}\delta(nT-T)\), where \(T\) is the sampling duration and \(0\leq \rho  \leq 1\). The results are also depicted in Fig. \ref{fig2}, where \(\rho = 0.3\). It has been demonstrated that DM-CSS provides the best performance relative to other counterparts. An important aspect to highlight here is that the performance of ePSK-LoRa(\(2,4\)) which is similar to that of DM-CSS in an ideal AWGN channel, deteriorates considerably in the fading channel.
\subsection{Bit Error Performance under Phase Offset}
\begin{figure}[tb]\centering
\includegraphics[trim={52 0 0 0},clip,scale=0.86]{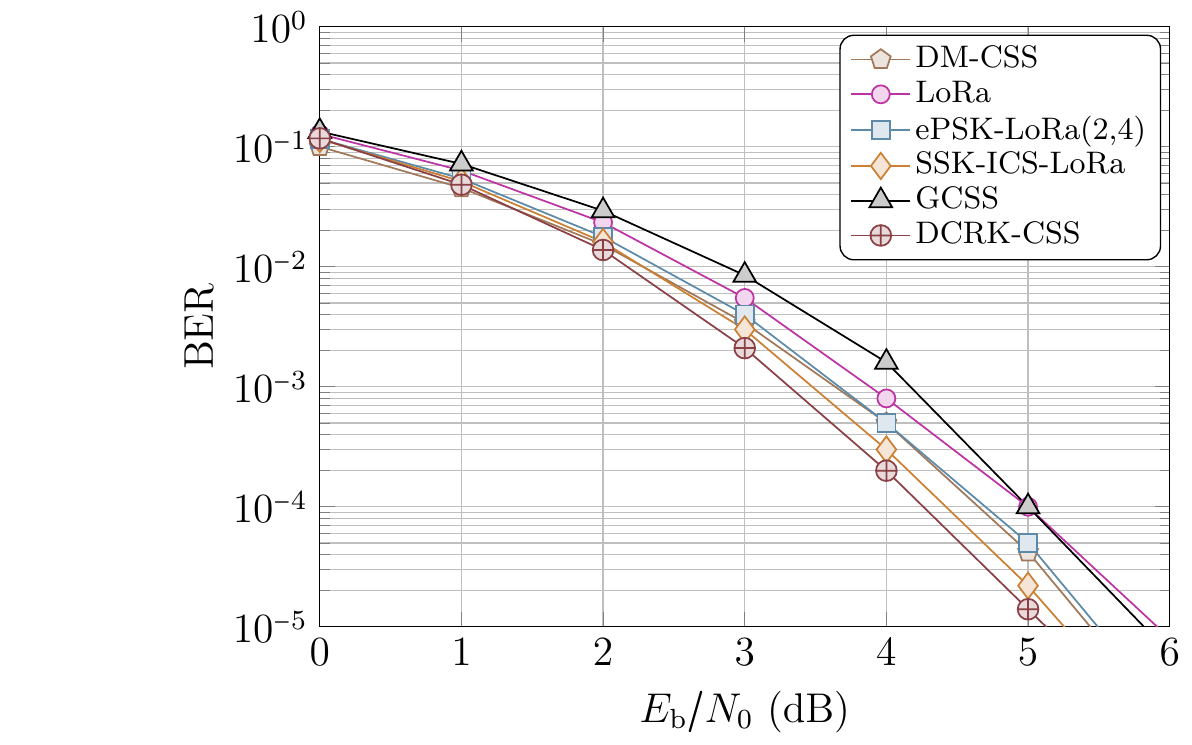}
  \caption{BER performance for different schemes considering non-coherent detection, \(\psi=\sfrac{\pi}{8}\), AWGN channel, and \(\lambda = 9\).}
\label{fig3}
\end{figure}

In this section, we analyze the performance of DM-CSS considering PO, which is expected to exist in low-cost devices. To this end, the received symbol corrupted by PO and AWGN is given as \(\exp\{j\psi\} s(n) + w(n)\), where \(\psi\) is the PO, and \(w(n)\) are the samples of AWGN. Considering non-coherent detection, PO of \(\psi=\sfrac{\pi}{8}\) and \(\lambda = 9\), the BER performance is presented in Fig. \ref{fig3}. We observe that the performance of DCRK-CSS, SSK-ICS-LoRa and LoRa are most robust because they do not incorporate any information in the PS of the transmit symbol. On the other hand, DM-CSS, ePSK-LoRa\((2,4)\) and GCSS suffer from the PO because either multiple chirps are simultaneously multiplexed, or the symbols carry additional information in the PS of the transmit symbols. Nonetheless, the performance of DM-CSS is still acceptable, bearing in mind that DM-CSS transmits maximum number of bits, compared to other schemes. A BER penalty of \(0.6\) dB is observed for DM-CSS considering PO of \(\psi=\sfrac{\pi}{8}\) relative to when \(\psi=0\).
\subsection{Bit Error Performance under Frequency Offset}
\begin{figure}[tb]\centering
\includegraphics[trim={52 0 0 0},clip,scale=0.86]{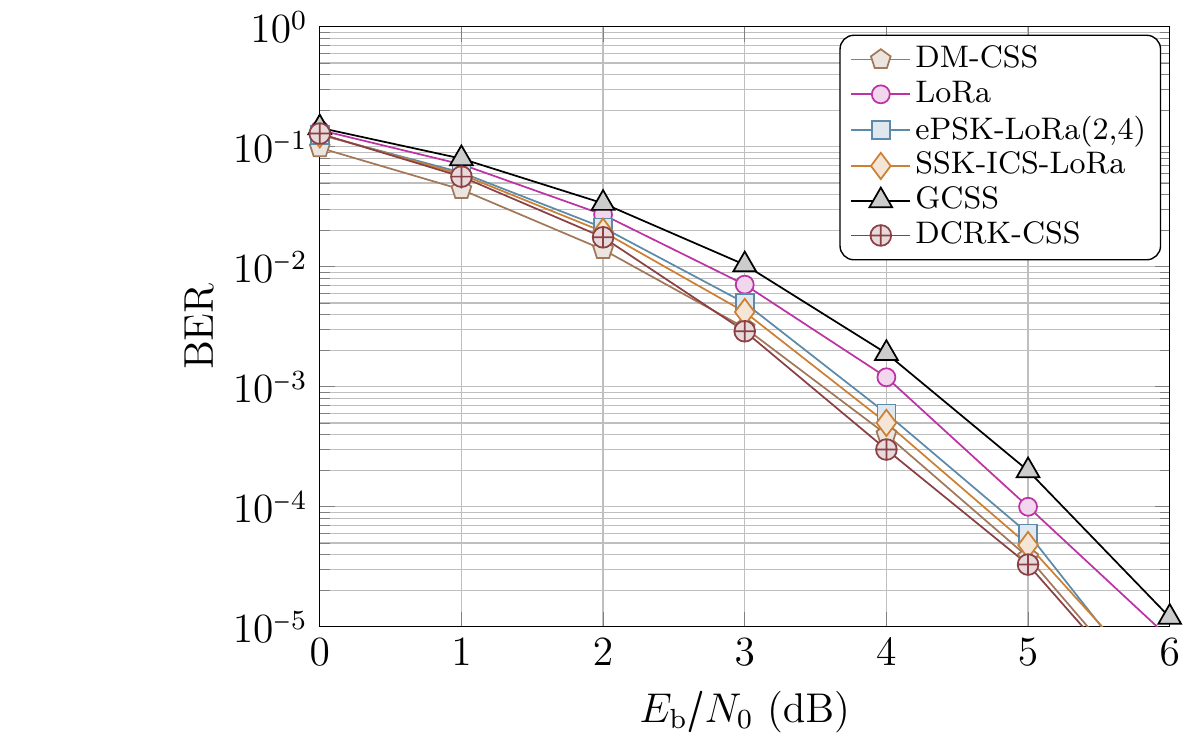}
  \caption{BER performance for different schemes considering non-coherent detection, \(\Delta f=0.1\), AWGN channel, and \(\lambda = 9\).}
\label{fig4}
\end{figure}

The BER performance of DM-CSS considering the influence of carrier FO is provided in Fig. \ref{fig4}. The received symbol incorporating the impact of FO is \(\exp\{\sfrac{j2\pi \Delta f n}{N}\} s(n) + w(n)\). Carrier FO accumulates phase rotation sample-by-sample in the symbol. To attain Fig. \ref{fig4}, we use \(\Delta f = 0.1\) and \(\lambda =9\).  We observe that the performance of DM-CSS is marginally better than ePSK-LoRa(\(2,4\)) and SSK-ICS-LoRa. DCRK-CSS exhibits the best performance, whereas, the BER performance of LoRa and GCSS considering the influence of FO is agnostic. 
\vspace{-2mm}
\subsection{Advantages and Disadvantages of DM-CSS}
Some observable advantages of DM-CSS are: (i) higher achievable SE over other alternatives; (ii) improved EE relative to state-of-the-art schemes; (iii) the impact of FO is the lowest for DM-CSS among DM-CSS, ePSK-LoRa(\(2,4\)) and GCSS; and (iv) DM-CSS provides the best performance in frequency-selective \(2\)-tap fading channels. Nonetheless, there are some distinctive disadvantages as well such as: (i) the performance of DM-CSS, ePSK-LoRa(\(2,4\)) and GCSS is negatively impacted under the influence of PO and FO (roughly \(0.6\) dB and \(0.8\) dB considering (\(\psi=\sfrac{\pi}{8}\)) and (\(\Delta f = 0.1\)), respectively), whereas the performance of LoRa, SSK-ICS-LoRa, DCRK-CSS is somewhat agnostic; (ii)  DM-CSS does not possess the constant envelope property, therefore, peak-to-average power ratio (PAPR) could be a concern at some stage of system realization; and (iii) the detector involves two ML criteria to decode \(\hat{\alpha}_\mathrm{e}\) and \(\hat{\alpha}_\mathrm{o}\), that adds complexity to the receiver design.

It is highlighted that the primary reason for BER degradation of schemes under PO and FO is either because multiple chirps are multiplexed to increase the achievable rate or due to additional information being transmitted via the PS of the chirped symbol. However, the schemes that transmit only one chirp, such as LoRa, SSK-ICS-LoRa, DCRK-CSS, achieve relatively lower bit rates. Anyhow, the superiority of DM-CSS is also evident from the fact that it is reasonably resilient against FO and PO and attains the highest achievable SE among the considered approaches. 
\vspace{-2mm}
\section{Conclusions}
In this letter, we proposed DM-CSS that is capable of substantially improving the SE relative to other state-of-the-art alternatives. A non-coherent detection process is succinctly explained. The proposed scheme was shown to outperform LoRa, SSK-ICS-LoRa, GCSS and DCRK-CSS in terms of EE/BER over an AWGN channel for different \(\lambda\). The performance of DM-CSS was found to be among the best in a fading channel.  We also derived mathematical formulations to prove the orthogonality of DM-CSS symbols. Furthermore, the resilience of DM-CSS against PO and FO impairments have been demonstrated. It is foreseen that the achievable SE of DM-CSS can be enhanced using different chirp rates as in DCRK-CSS. Such design is left out as a possible topic for future research.
\bibliographystyle{unsrt}
\bibliography{biblio}
\end{document}